# Simultaneous imaging of voltage and current density of flowing electrons in two dimensions

L. Ella[1], A. Rozen[1], J. Birkbeck[2,3], M. Ben-Shalom[2,3,4], D. Perello[2,3], J. Zultak[2,3], T. Taniguchi[5], K. Watanabe[5], A.K. Geim[2,3], S. Ilani[1], and J.A. Sulpizio[1*]

[1] *Department of Condensed Matter Physics, Weizmann Institute of Science, Rehovot 76100, Israel.*
[2] *School of Physics & Astronomy, University of Manchester, Manchester M13 9PL, United Kingdom.*
[3] *National Graphene Institute, University of Manchester, Manchester M13 9PL, United Kingdom.*
[4] *Department of Physics, Tel-Aviv University, Israel.*
[5] *National Institute for Materials Science, 1- 1 Namiki, Tsukuba, 305- 0044 Japan.*
[*] Correspondence to: joseph.sulpizio@weizmann.ac.il

**Electron transport in nanoscale devices can often result in nontrivial spatial patterns of voltage and current that reflect a variety of physical phenomena, particularly in nonlocal transport regimes. While numerous techniques have been devised to image electron flows, the need remains for a nanoscale probe capable of simultaneously imaging current and voltage distributions with high sensitivity and minimal invasiveness, in magnetic field, across a broad range of temperatures, and beneath an insulating surface. Here we present such a technique for spatially mapping electron flows based on a nanotube single-electron transistor, which achieves high sensitivity for both voltage and current imaging. In a series of experiments using high-mobility graphene devices, we demonstrate the ability of our technique to visualize local aspects of intrinsically nonlocal transport, as in ballistic flows, which are not easily resolvable via existing methods. This technique should both aid in understanding the physics of two-dimensional electronic devices, as well as enable new classes of experiments that image electron flow through buried nanostructures in the quantum and interaction-dominated regimes.**

Over the last few decades, a growing variety of materials and devices have emerged with electron flow described outside the framework of semi-classical diffusive transport. In these systems, electrons propagate either ballistically through the bulk[1], ballistically via edge transport, such as in topological insulators[2,3], or most recently - collectively via hydrodynamic flow arising from strong electron-electron interactions[4]. In such non-diffusive flows, the relation between current and voltage is no longer local, and the spatial structure of both the current and voltage can take non-trivial forms.

Our understanding of electron flow in two-dimensional systems predominantly stems from transport measurements on lithographically defined devices, which are an indispensable tool in studying many aspects of non-diffusive flows. Such measurements sample the electrochemical potential only at discrete points in space, typically along the edge of a device. However, as noted long ago by Landauer[5], such measurements also have inherent limitations: since transport measurements use fixed voltage probes that sample the potential only at discrete spatial points, they cannot provide a full map of the current and voltage distribution. Furthermore, in the ballistic regime where the electron mean free path is large compared to the probe and device dimensions, such fixed probes sample the electronic distribution in a manner that depends on their geometric details, and often disrupt the flow they are aiming to measure[1,6]. These limitations illustrate the need for a non-invasive local probe that can provide real-space visualization of the fundamental properties of electronic transport; namely the electrostatic potential and current density, in real-space, and particularly in regimes where some combination of quantum, ballistic, and electron-electron interaction effects become dominant.

Several scanning probe techniques have been previously developed that separately image voltage[7–10] or current[11–15] in two-dimensional systems, each with specific advantages, though the emerging classes of materials can benefit from a probe with a broader scope of measurement capabilities. A highly desirable feature for a voltage imaging tool is the ability to image electrons buried beneath insulating surfaces (currently achieved by Kelvin probe[10] and optical methods[8,9]), due to their increasing prevalence for improving mobility. However, if such a probe aims to address delicate low-energy physics phenomena, it must also possess high voltage sensitivity (currently obtained only with STM potentiometry[7]). Furthermore, it must be able to image across a wide temperature range to capture a broad array of phenomena, and do so non-invasively at small carrier densities and energy scales. Along with the voltage, one would ideally like to image simultaneously the local current density flowing through a device as well. Several techniques have recently excelled in imaging current by measuring the magnetic field it produces (scanning SQUIDS[11,13,14,16] and NV centers[15]), although they are limited to operate under small externally applied magnetic fields. Additional scanning techniques (scanning gate microscopy[17], NSOM[18,19], photocurrent[20], optical[21,22] and microwave impedance[23,24]) have proven crucial for visualizing other aspects of transport through devices. A promising candidate for imaging the properties of flowing electrons is the scanning single electron transistor (SET)[25–27], owing to its extreme voltage sensitivity. In the past, SETs have been used primarily for imaging equilibrium properties (e.g. workfunction and electronic compressibility) and for resolving questions about the spatial distribution of quantum hall edge states[28]. Their possible capacity to image voltage drops and current density of flowing electrons has remained unexplored.

In this work, we demonstrate a new technique employing nanotube-based SETs[29] to simultaneously image the voltage drop and the current distribution of flowing electrons in two dimensions. The technique has nanoscale spatial resolution and microvolt voltage sensitivity, operates from cryogenic temperatures up to room temperature and at large magnetic fields, and is minimally invasive to the flow, making it especially suitable for visualizing non-diffusive transport. We benchmark the technique by mapping electron flow in high-mobility graphene/hBN devices[30]. We observe the evolution from diffusive flow, in which electrostatic potential falls gradually along the device, to ballistic flow, where it drops sharply at its contacts. Independently, we show how the SET can be used to image current streamlines, and how such voltage and current maps can reveal detailed information about electron flow within the bulk of a device, which is difficult to obtain via conventional methods. The demonstrated technique paves the way to imaging non-diffusive flow phenomena in a variety of buried nanostructures created from a growing list of novel materials and devices.

The principle behind our measurement technique is shown schematically in figure 1. In a typical transport measurement (fig. 1a), fixed, lithographically defined contact electrodes patterned along the perimeter of a device are used to measure the *electrochemical* voltage drop in response to a flowing current, yielding both the longitudinal ($\rho_{xx}$) and Hall ($\rho_{xy}$) resistivity. While these quantities are well-defined locally for diffusive transport, in ballistic flow the contact electrodes sample only the electrons with momenta directed toward them[1,6], yielding an averaged electrochemical potential that depends on the contact orientation and precise geometry. Additionally, such contacts emit thermalized electrons, thereby randomizing their direction, and thus can interfere with the flow. Our technique replaces the fixed voltage probes with a scanning SET, which can non-invasively sample the local, out-of-equilibrium *electrostatic* potential anywhere in space. The SET couples capacitively to the region in the sample above which it is scanning, such that the local potential of the sample strongly modulates the current passing through it[25–27]. By monitoring the SET current as it scans across the sample under study, we can thus isolate the electrostatic potential, $\delta\phi$, generated in response to an applied total AC current, $\delta I$ (fig. 1b). The measured quantity, $\delta\phi/\delta I$, has units of resistance and captures the local change in electrostatic potential (voltage drop) due to the flowing electrons[6,31].

Analogous to $\rho_{xx}$ and $\rho_{xy}$ conventionally measured in transport, the quantity $\delta\phi/\delta I$ allows us to map in real space two independent properties of a device under study: If measured at zero magnetic field, it gives the voltage drop associated with varying longitudinal resistivity, dropping more sharply in locations that are more resistive (fig. 1c). Under the application of a weak perpendicular magnetic field $\pm B$, we independently resolve the Hall voltage associated with the flow, $\delta\phi_H/\delta I \equiv (\delta\phi_{+B} - \delta\phi_{-B})/2\delta I$ (fig. 1d). Since the difference in Hall voltage between two spatial points separated by $\Delta y$ is directly related to the current passing between them via $j\Delta y = \frac{ne}{B} \cdot \Delta(\delta\phi_H)$ (where $j$ and $n$ are the local current and carrier densities and $e$ is the electron charge), spatially resolved measurement of $\delta\phi_H/\delta I$ directly yields a map of the local current density.

Having outlined the basis for our technique, we now demonstrate its ability to image the voltage drop of flowing electrons using a high-mobility graphene/hBN device at $T = 4K$. The graphene is patterned into a channel 11μm wide ($W$) and 25μm long ($L$) (fig. 2a), with source and drain contacts exhibiting a typical 2-point resistance that is sharply peaked as the carrier density is tuned near charge neutrality via the back gate voltage (fig. 2b). The spatial map of the electrostatic potential of the flowing electrons taken at charge neutrality (fig. 2c) exhibits an overall linear drop along the channel, indicating diffusive behavior, though with clear local fluctuations reflecting disorder-induced resistivity variations. From the slope of this voltage drop, we find the transport mean free path $l_{\text{tr}} = \frac{h}{2e^2 k_F W} \cdot \Delta x / \Delta(\frac{\delta \phi}{\delta I})$, to be 0.9μm, which, as expected for diffusive transport, is much smaller than the device dimensions ($h$ is Planck's constant and $k_F$ is the Fermi wavelength, which we estimate from the residual density fluctuations at charge neutrality). Similar imaging of diffusive flow has been done previously using Kelvin probe microscopy under the application of a high bias voltage[32] (~2 volts). The extreme sensitivity of our scanning SET allows us to perform these measurements with three orders of magnitude smaller bias (2.5mV) which combined with higher mobility samples and cryogenic temperatures, allows us to access the regime of ballistic flow. Indeed, when we image the voltage drop for a hole density of $n = 1 \cdot 10^{12} \text{cm}^{-2}$ we measure a strikingly different map that strongly resembles the textbook picture of ballistic transport[6] (fig. 2d). Here, the resistance is localized almost entirely at the graphene-contact interface, as visualized by the sharp, step-like voltage drops. In the bulk of the device, the electrostatic potential is nearly flat, with only a small, residual drop due to the large, but finite mean free path $l_{\text{tr}} = 26$μm.

The spatial maps of $\delta \phi / \delta I$ allow us to quantitatively distinguish the different contributions to the total resistance of the channel. Figure 3a shows the electrostatic potential measured along the center of the device (dashed line, fig. 2a) at three carrier densities for both holes and electrons. Within the bulk of the device, the potential drop is highly electron-hole symmetric (fig. 3a, between vertical dashed lines), with resistivity decreasing as density is increased. The contact resistance (fig. 3a, vertical arrow), however, exhibits a marked electron-hole asymmetry[33]. This is shown systematically in figs. 3b and 3c, where the extracted bulk conductivity, $\sigma_{xx}$, and contact resistance, $R_c$, are plotted as a function of the carrier density. The larger contact resistance for holes most likely arises from the formation of p-n junctions between the hole-doped graphene bulk and electron-doped regions near the contacts due to differences in the workfunction. For electron doping, where no p-n junctions should form, we can compare against the Sharvin resistance[34] of an ideal contact. We find that the imaged $R_c$ at high electron density approaches the predicted Sharvin value for 4-fold spin/valley degenerate fermions in graphene, $R_{\text{sharvin}} = \frac{h\pi}{4e^2 W k_F}$ (fig 3d, dashed line), to within a factor of two, indicating large contact transparency (up to $T\sim0.5$, top inset Fig 3c). In Supplementary S4 we show an additional example in which we image electron flow around an obstacle, revealing electrostatic potential maps within the bulk of a sample that would be challenging to obtain with existing techniques.

We now turn to imaging of the local current distribution, which we accomplish via mapping the Hall voltage in real space. We use a second graphene device with a bend geometry (Fig. 4a). In this case, we expect the Hall voltage to reflect the fact that the current follows the bend. Figures 4b and 4c show these voltages imaged at small positive and negative perpendicular magnetic fields $B = \pm 20$mT. Notably, the equipotential lines are now tilted with respect to the channel direction, due to the addition of the Hall voltage to the longitudinal voltage drop. By subtracting these two voltage maps, we remove the longitudinal resistance, which is symmetric in $B$, isolating the contribution of the Hall voltage, $\delta\phi_H/\delta I$, normalized by the current. To interpret this Hall voltage map, we first note that the density across the channel is nearly constant. We determine this by imaging with the SET the local density of states along the y-axis of the graphene device (fig 4d), and observing from the variation in the charge neutrality point that the spatial fluctuation in charge density is at most $\delta n_{disorder} \approx 3 \cdot 10^9 \text{cm}^{-2}$, a small fraction of the average density, $n = -1.1 \times 10^{11} \text{cm}^{-2}$. If local density variations were present, they could be similarly measured by the SET, making the technique applicable to non-uniform devices as well. Since the Hall electric field must locally balance the Lorentz force, the Hall voltage map, together with the density, allows us to directly obtain the electron flow streamline function[1,35], $\delta\psi = \frac{ne}{B} \cdot \delta\phi_H$, whose derivative gives the local current density, $j = \hat{z} \times$ ($\nabla\delta\psi$). In Fig. 4e we plot the resulting streamlines (accumulated current is labeled) superimposed on the independently measured zero field equipotential contours, showing that the current streamlines clearly snake around the bend in the device. We can additionally determine the total current through the device using our imaging technique, independently of transport measurements via $\delta I_{imaged} = \delta\psi(y=W) - \delta\psi(0)$, where $y$ is the coordinate along the channel profile. Indeed, we find that $\delta I_{imaged} = 3.4 \pm 0.2\mu A$ matches well the total current $\delta I = 3.2\mu A$ measured via transport with no free parameters. The smearing of the streamlines near the channel edges is related to the finite spatial resolution, which is limited by the probe-sample separation in this specific measurement (SI 1).

Examining the electron flow more closely, we note a sharp *increase* in resistance at the entrance to the bend, as indicated by the bunching together of the equipotential contours. Diffusive flow with homogenous bulk conductivity should in fact show a resistance *decrease* at the bend, since there the channel width increases. This is indeed predicted by a simulation of diffusive flow for this exact device geometry (fig. 4f). For ballistic flow, on the other hand, a change of channel width acts as a reflecting barrier, leading to a resistance increase. An electron billiards simulation of ballistic flow with diffusive boundaries for this device geometry (fig. 4g) shows a clear bunching of the equipotential contours around the bend as in the experiment, albeit with a somewhat reduced resistance increase. The remaining discrepancy is likely due to additional disorder and mechanical stresses near the etched boundaries of the bend. These detailed maps of both the voltage and current around the bend highlight the advantage of our technique for visualizing the flow inside the bulk of the sample, which is otherwise challenging to obtain via existing techniques.

Having demonstrated the utility of our technique with a few representative examples, we now discuss its broad range of applicability and benchmark its performance. In the above measurements, we achieved a voltage sensitivity of $\sim 2\mu V/\sqrt{Hz}$ (SI 3). While these measurements were performed at $T = 4K$, we have imaged flows from cryogenic temperatures all the way to room temperature (SI 4,5). The technique also functions over a large range of magnetic fields (>10T), as transport through the SET is only weakly field dependent. We also stress that because no charge is directly transferred between the SET and the sample, buried structures impose no measurement restrictions. Our measurements reached a spatial resolution of ~100nm (SI 1), limited by the lithographic dimension of the SET and its height above the sample. In principle, these values may be scaled down to obtain a resolution in the range of a few tens of nanometers. Moreover, in contrast to tip-based capacitive probes, our SET is embedded in a planar geometry surrounded by screening electrodes, resulting in an exponentially-localized point spread function (SI 1) which is crucial for resolving fine local features within a large background. We characterize the invasiveness of the SET by measuring how strongly it gates the sample under study, and find the induced local density variation to be $\delta n_{invasiveness} \approx 1 \cdot 10^9 \text{cm}^{-2}$, which is smaller than the influence of disorder (SI 2), making its effect negligible.

The current imaging sensitivity is directly related to the voltage sensitivity via the Hall resistivity. While the SET can operate at any magnetic field, imaging unperturbed ballistic current requires the use of only a small probing field that does not affect the electron trajectories, namely, $W \ll r_c$ (where $r_c = \hbar\sqrt{\pi n}/eB$ is the cyclotron radius). The resulting sensitivity in measurements of the current density is $\sim \frac{10\text{nA}/\mu\text{m}}{\sqrt{\text{Hz}}} \cdot \sqrt{\frac{n}{10^{11}\text{cm}^{-2}}}$. For metallic samples (e.g. $n = 10^{15}\text{cm}^{-2}$) this is comparable to the sensitivity of the present scanning SQUIDs and NV centers. For typical densities in semiconductors and semimetals (e.g. $n = 10^{11}\text{cm}^{-2}$), though, our sensitivity is better by two orders of magnitude. Since our current imaging technique relies on Hall voltages, we are restricted from imaging supercurrents due to the compensating flow of holes along with the electrons. However, because our technique does not require deconvolution of an imaged magnetic field that can be highly nonlocal, we can otherwise easily obtain quantitative maps of current flows, especially when the current has a non-trivial profile within channels, making it very useful for deciphering non-diffusive flow regimes.

In summary, we have introduced a new scanning SET-based method for visualizing both the voltage drop and current density of flowing electrons in two-dimensional systems, with high sensitivity and minimal invasiveness. Applying this technique to high-mobilty graphene/hBN devices, we are able to spatially image the voltage drop of flowing electrons spanning from diffusive to ballistic flow. With the addition of a weak magnetic field, we further image the local current density, thus performing a complete local characterization of electron flow in a ballistic system. This technique holds promise for imaging an array of phenomena presently under intense focus, such as hydrodynamic electron flow, Dirac electron optics, and Andreev reflections in magnetic field. In a more practical direction, the technique is well suited for fully characterizing transport effects due to disorder and other scattering mechanisms in a broad class of novel materials and devices, where imaging the local electronic flow patterns can provide crucial information that is otherwise inaccessible via conventional methods.

## Methods

**Device fabrication:** Scanning SET devices were fabricated using our nano-assembly technique, presented in detail in Ref [29]. The graphene/hBN devices were fabricated using electron-beam lithography and standard microfabrication procedures[30].

**Measurements:** The measurements are performed in a home-built, variable temperature, Attocube-based scanning probe microscope. The microscope operates in vacuum inside a liquid helium dewar with a superconducting magnet, as well as under ambient conditions. The measurement apparatus is mechanically stabilized using Newport laminar flow isolators. For variable temperature studies, a local resistive heater is used to heat the sample under study from T=4K up to room temperature. A DT-670 diode thermometer chip is mounted proximal to the sample and on the same printed circuit board for precise temperature control. Voltages and currents (for both the SET and sample under study) are sourced using a home-built DAC array, and measured using a home-built, software-based audio-frequency lock-in amplifier consisting of a Femto DPLCA-200 current amplifier and NI-9239 ADC. The local gate voltage of the SET is dynamically adjusted via custom feedback electronics employing a least squares regression algorithm to prevent disruption of the SET's working point during scanning and ensure reliable measurements.

**Data availability:** The data that support the plots and other analysis in this work are available from the corresponding author upon request.

in Viscous Electronics. *Phys. Rev. Lett.* **119,** (2017).

**Acknowledgements:** We thank G. Falkovich, L. Levitov, A. Shytov and A. Stern for discussions and D. Mahalu for electron-beam lithography. We further acknowledge support from the Helmsley Charitable Trust grant, the ISF (grant # 712539), WIS-UK collaboration grant, and the ERC-Cog (See-1D-Qmatter, # 647413).

**Author Contributions:** LE, AR, SI, and JAS created the SETs, performed the measurements, and analyzed the data. JB, DP, JZ and MBS fabricated the graphene devices. KW and TT supplied the hBN crystals. LE, SI, and JAS wrote the manuscript with input from all authors.

**Competing financial interests:**

The authors declare no competing financial interests.

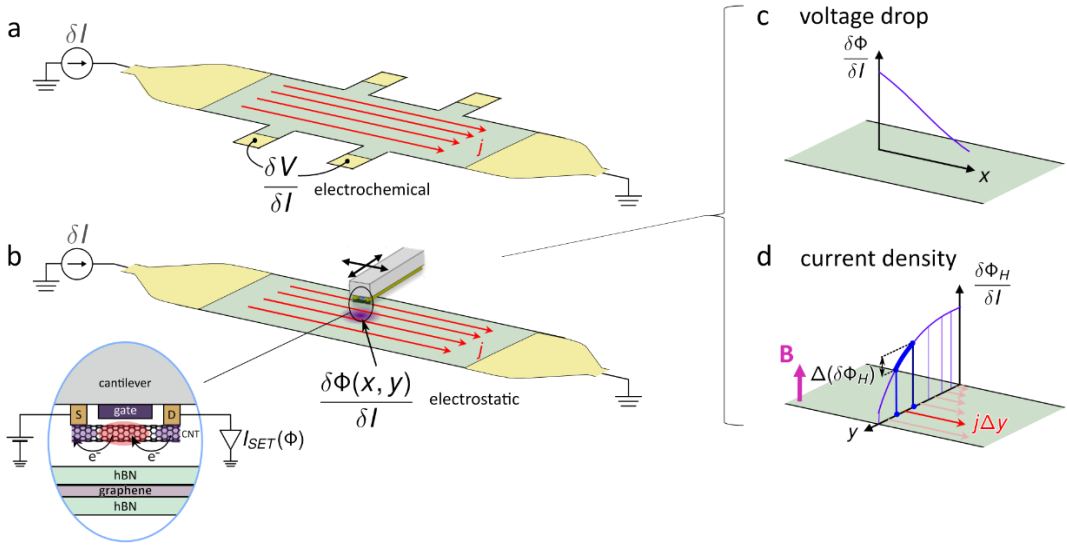

**Figure 1. Overview of the nanoscale voltage and current imaging technique. a**. Conventional transport measurement; AC current $\delta I$ (current density $j$, red arrows) is passed between the outer contacts (yellow) of a hall bar device (green). The *electrochemical* potential difference ($\delta V$) between pairs of contacts at the edge of the hall bar is measured to infer the resistivity and Hall voltage at fixed spatial positions. **b.** Illustration of SET-based imaging technique. In contrast to the fixed position data obtained by conventional transport measurements, a nanotube-based SET (inset) positioned at the end of a scanning probe cantilever is rastered (black arrows) to image the *electrostatic* potential (voltage drop), $\delta\phi(x,y)$, at every spatial position within a device generated by the flowing electrons with total AC current $\delta I$. Inset: Side view of the SET and schematic of the devices imaged in this paper - hBN/Graphene heterostructures atop Si/SiO$_2$. The SET consists of a pristine carbon nanotube, nano-assembled[29] on top of source and drain contacts (yellow) and suspended above a gate electrode (purple), positioned near the edge of the cantilever. At cryogenic temperatures, a quantum dot forms in the suspended segment (red) separated by p-n junctions from hole-doped nanotube leads (blue). The current flowing through the nanotube (black arrows) depends strongly on the electrostatic potential induced by the sample via Coulomb blockade physics, allowing measurement of the local voltage with extreme sensitivity. Note that during operation, no electrons are transferred between the SET and the device under study, and no mechanical force is applied, enabling the measurement of buried electronic devices with minimum disturbance. The technique allows simultaneous imaging of two basic quantities **c.** Voltage drop of flowing electrons $\delta\phi/\delta I$ (normalized by total current), obtained directly by measuring the local electrostatic potential in sync with the flowing current . While the figure illustrates the drop along the channel, in practice we measure the voltage drop in both spatial directions. **d**. Current density: Adding a small perpendicular magnetic field, $\pm B$ (purple) generates a local Hall voltage, $\delta\phi_H(x,y)$ (see main text for definition), without modifying the flow pattern of the current. By measuring the Hall voltage that drops between any two points in the sample, $\Delta(\delta\phi_H)$, (e.g. blue points in sketch), together with the local density, $n(x,y)$, we directly obtain the current flowing between these two points: $j\Delta y = \frac{ne}{B}\cdot\Delta(\delta\phi_H)$, where $e$ is the electron charge. More generically, we can get a full map of the local current density (magnitude and direction) via: $\vec{j} = \frac{ne}{B}\hat{z}\times\nabla(\delta\phi_H)$ (where $\hat{z}$ is a unit vector perpendicular to the plane).

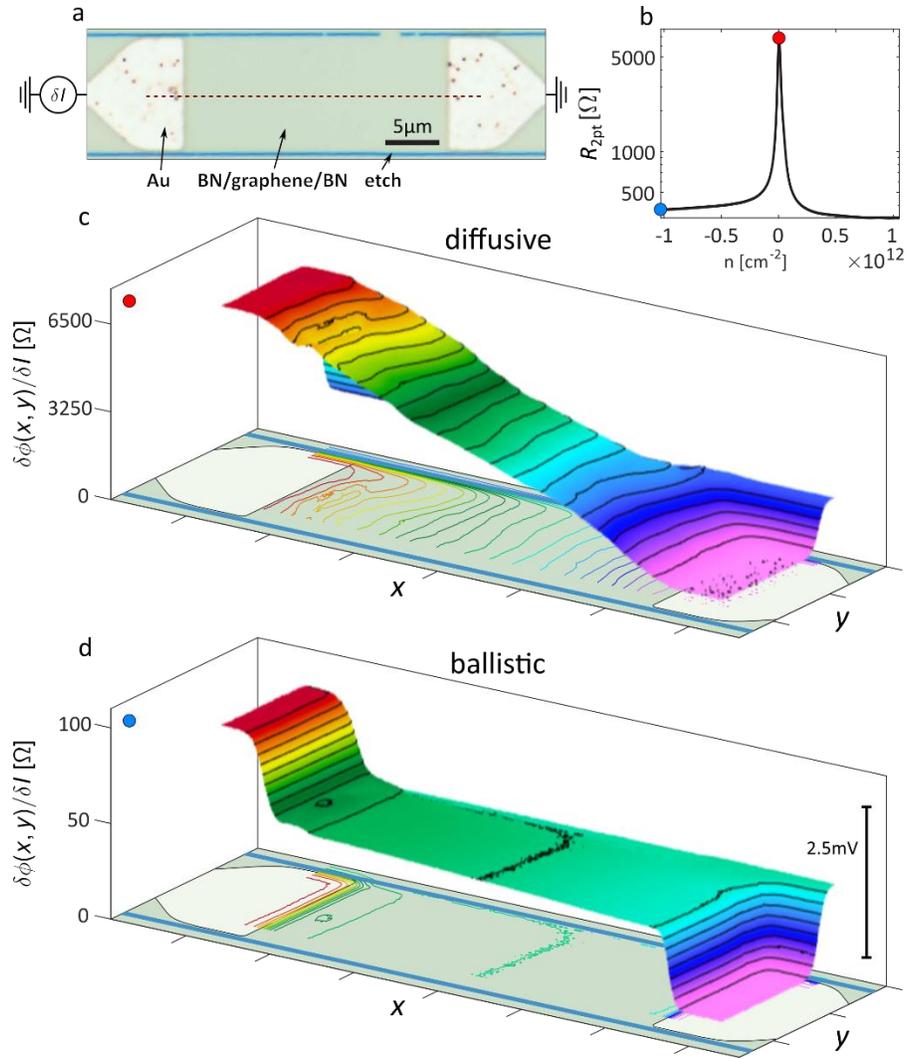

**Figure 2. Spatial imaging of the voltage drop of flowing electrons in the diffusive and ballistic regimes**. **a,** An optical image of the device, consisting of a conducting mesoscopic channel defined within a single-layer graphene/hBN sandwich (green), using chemically etched boundaries (blue). The current, $\delta I$, is passed between a pair of gold contacts (yellow). **b.** Two-probe resistance of the device at $T = 4K$, measured as a function of carrier density (tuned via back gate voltage). **c.** Diffusive transport imaged near charge neutrality (red dot, panel b) where the graphene resistivity across the bulk of the device is dominant. The imaged electrostatic potential is shown, normalized by the total current, $\delta\phi(x,y)/\delta I$ (units of resistance). The bottom plane shows the equipotential contours superimposed on the schematic of the graphene channel and contacts, indicating that the voltage drops gradually between the contacts, with some local deviations due to increased disorder near charge neutrality. **c.** Ballistic transport imaged at a hole density of $1 \cdot 10^{12} \text{cm}^{-2}$ (blue dot, panel b). The voltage drops in a step-like manner at the interface between the contacts and the graphene channel, and is rather flat across the bulk of the device. The total voltage applied across the device in this measurement is $\sim 2.5 mV$ (side bar).

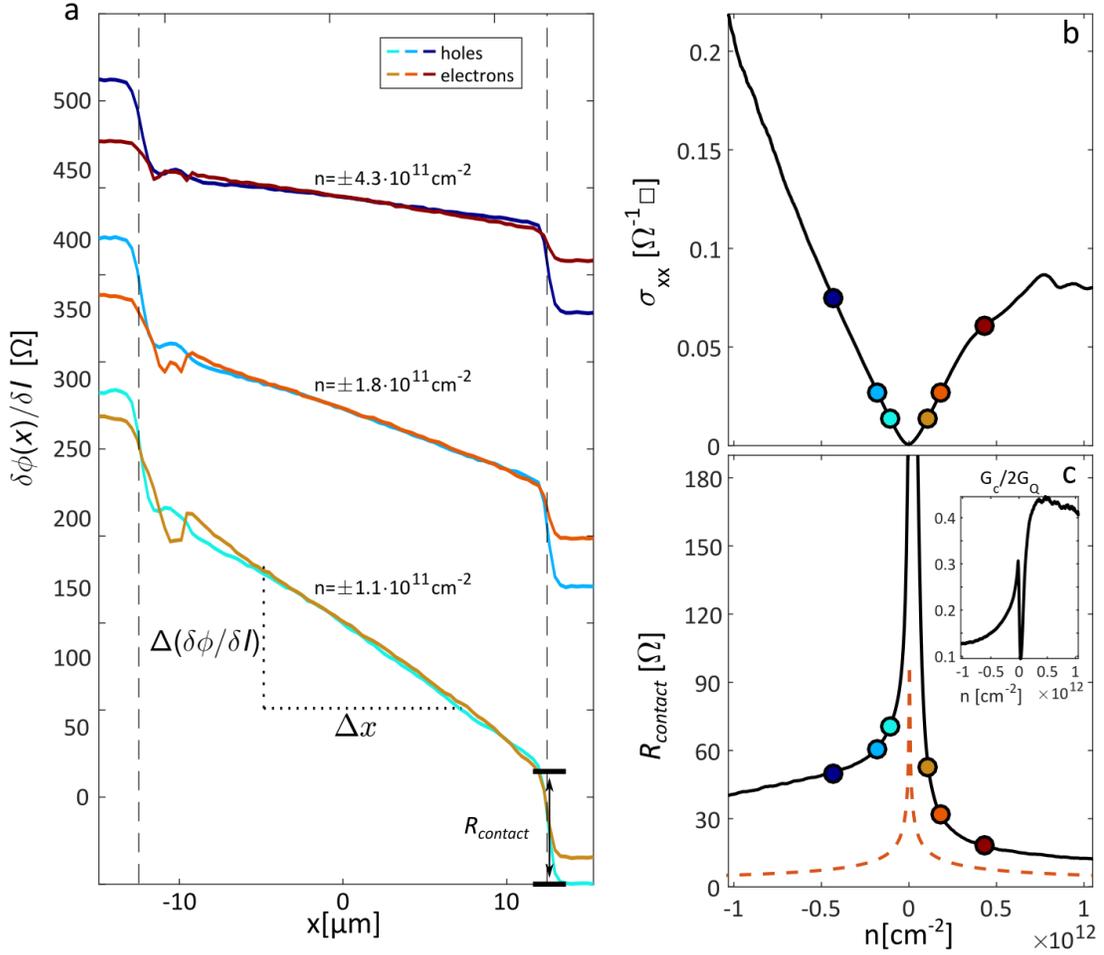

**Figure 3. Extracting local quantities from electrostatic potential maps. a.** Line traces of the imaged voltage drop of the flowing electrons, $\delta\phi/\delta I$, as a function of the spatial coordinate along the center of the channel, $x$ (dashed line, in fig. 2a). Measurements are taken at three pairs of equal electron/hole carrier densities. A vertical offset between equal density pairs is added for clarity. The bulk graphene resistivity is determined directly from the local slope of the voltage, normalized by the width of the channel ($W$) via $\rho_{xx} = \Delta(\delta\phi/\delta I)/\Delta x \cdot W$. Notably, at the same carrier density the measured local slopes for electrons and holes are similar. The contact resistance $R_C$ is defined as the size of the $\delta\phi/\delta I$ step between the contacts and the onset of the bulk voltage slope (indicated by black bars), and shows a large degree of electron/hold asymmetry. Near the left contact there is a localized voltage artifact that resulted from voltage cycling of the back gate, which locally disrupted the operation of the SET. **b., c.** Graphene conductivity $\sigma_{xx}$ and contact resistance extracted from electrostatic potential traces as in panel a, but measured as a continuous function of carrier density. Colored circles correspond to line cuts in panel a. The orange dashed line in panel c corresponds to the predicted Sharvin contact resistance limit. Inset: deduced contact transparency.

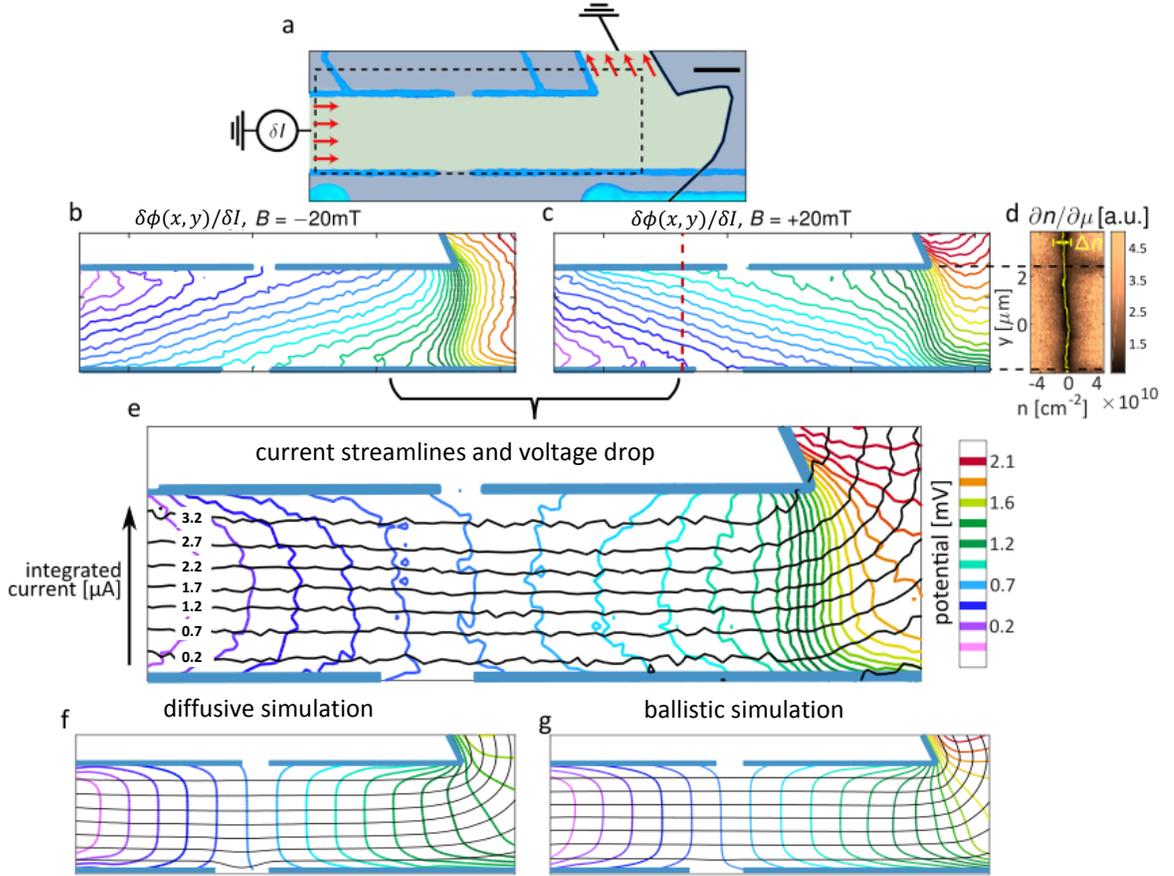

**Figure 4. Imaging the local current density in a graphene device with a bend. a.** Optical image of the device. The relevant channel is bounded by etched lines (blue) and a natural edge of the graphene (black). Irrelevant parts of the device are grayed out. The contact electrodes (beyond the field of view) inject AC current $\delta I$ at the left (red arrows) and collect it at the top (red arrows) around the bend. Scale bar is 2.5 μm. Dashed rectangle outlines the region imaged with the SET. **b.** and **c.** Equipotential contours of flowing electrons at small perpendicular magnetic fields, $B = \pm 20$mT, taken at a hole density of $n = 8.3 \times 10^{10}$cm$^{-2}$ set by the back-gate voltage. The magnetic field rotates the equipotential contours, allowing direct visualization of the local Hall angle. **d.** Electronic compressibility, $\partial n/\partial \mu$, measured along a line cut perpendicular to the channel (dashed line in panel c) as a function of the global density tuned by the back gate voltage. The dark feature near zero global density corresponds to the local charge neutrality point, and its variance (yellow line) is $\delta n_{disorder} = 3 \times 10^9$cm$^{-2}$, only 3% of the total density, including near the channel edge, justifying the use of a constant density in the subsequent analysis. **e.** Imaged current streamlines $\psi$ (black iso-contours) superimposed on the zero field voltage contours (color). The streamlines are normalized by the Hall resistance $R_H = \frac{B}{ne}$, such that $\delta\psi(x,y) = \frac{\delta\phi_{+B}(x,y) - \delta\phi_{-B}(x,y)}{2R_H}$, which we obtain directly from the difference of the maps at positive and negative fields (panels **b** and **c**). The local current is given by $\vec{j} = \hat{z} \times \nabla(\delta\psi)$. The numbers on the black streamlines count the total integrated current from the bottom of the channel, with 0.5μA spacing between the lines. **f.** and **g.** Voltage drop and current streamlines from simulations in the diffusive regime ($\vec{j} = \sigma\vec{E}$, constant $\sigma$) and the ballistic regime (billiard ball simulation).

# Supplementary Materials

**Simultaneous imaging of voltage and current density**

**of flowing electrons in two dimensions**

S1. Spatial resolution of imaging

S2. Invasiveness

S3. Voltage sensitivity

S4. Imaging electron flow across a resistive obstacle

S5. Imaging at elevated temperature and in ambient conditions

## S1. Spatial resolution of SET imaging

The spatial resolution of our technique is characterized by measurement of the point spread function (PSF) of the SET probe. Two factors contribute to the width of the PSF, which we denote as $\sigma$. The first is the intrinsic probe size, which is set by the size of the quantum dot formed in the suspended nanotube (NT). Since the NT is a long thin wire, its contribution $\sigma_i$ is highly anisotropic. An upper bound on the size of the quantum dot in the suspended NT is given by the lithographic spacing between its source and drain contact electrodes. Figs. S1a-c show a sequence of SEM images of the scanning SET probe, including a zoom on its active area. The actual size of the dot is often significantly smaller than the entire suspended NT segment, though, due to both the narrow width of the local gate as well as the finite size of the pn junctions that form the dot's tunnel barriers[1] In the current experiment $\sigma_i$ is ~100nm, but in principle it can be significantly reduced by lithographically defining a smaller structure. The magnitude of the component of $\sigma_i$ in the direction perpendicular to the NT is set by its diameter, which $is$ ~1$nm$. The second contribution to the width of the PSF, $\sigma_{\Delta z}$, arises from the probe-sample separation $\Delta z$ during the imaging, which is approximately linear in $\Delta z$ for $\Delta z \gg \sigma_i$, as we show below.

To best demonstrate the resolution, we use the NT SET to image another NT, which acts as a 'delta-function' spatial feature. The SET is scanned perpendicularly across the second NT with their axes parallel such that the $\sigma_i$ component is minimized to the $1nm$ NT diameter. We bias the sample NT to a voltage $V$, and image the electrostatic potential, $\phi$, that it produces as a function of the spatial coordinate. The measurement is shown in the fig S1d (red dots) together with a fit (black line) to a squared hyperbolic secant function. The full width half max (FWHM) of the fit is ~100$nm$, which gives the spatial resolution of the experiment.

For 2D imaging, we characterize the resolution by scanning the SET over a voltage step, which in the measurements below is realized by two independently biased graphene electrodes, separated by an etched line of width 150nm. Scanning across this voltage step (2D map, fig. S1e) shows a smeared voltage step (Fig. S1f, red dots) which is well described by the functional form $\sim \tanh(1.76x/\sigma_{\Delta z})$ (black). The derivative of this function (fig S1f, inset) is the *line-spread* function of the SET probe, which in this case differs from the point-spread (PSF) function by a few percent, and so we treat them as interchangeable[2]. Its FWHM, $\sigma_{\Delta z}$, gives our resolution. Repeating this experiment at different sample-probe separations (fig. S1g, dots) we find that $\sigma_{\Delta z}$ depends linearly on the ratio of capacitances between the SET and its local gate and the SET and the sample, $C_{local}/C_{sample}$ (schematically shown in Fig. S1h). In the measurements in this figure we reached a resolution of $350 nm$, while in other 2D experiments we have obtained resolutions down to $85 nm$ (section S5 below). The achieved resolution is not an intrinsic property of the SET, but simply reflects the closest approach distance in these the specific experiments.

A finite-element electrostatic simulation that takes into account the lithographic dimensions of the SET successfully reproduces the shape of the measured PSF as well as the linear relation between the FWHM of the PSF and the capacitance ratio $C_{local}/C_{sample}$ (black crosses, fig S1g). The ratio $C_{local}/C_{sample}$ gives a reliable predictor of our height over the sample, and is the quantity that we use in practice to navigate above the sample. Like in the experiment, the finite element modeling shows that this ratio scales linearly with the FWHM of the PSF (red line). The inset indicates that the capacitance ratio scales linearly with sample-probe separation, as expected.

It is important to note that the smearing of the imaged voltage steps fits extremely well to the $tanh$ function, implying that our PSF decays exponentially fast at large distances. This is in contrast to what is observed with other capacitive-based probes based on a sharp tip, including tip based scanning SETs[3], where the PSF decay has a long tail due to the range of the coulomb interactions. The reason for the highly local PSF in our case is that the NT SET is effectively surrounded by a plane of metallic electrodes (see Fig. S1a-c), which screen the field lines far away from the SET.

Similar to other capacitive based scanning probes, an important advantage of the scanning SET is that it can image the physics even at large probe sample separations (though with reduced resolution and voltage sensitivity). This is in contrast to techniques such as AFM and STM, in which the interaction between probe and sample is short-ranged, necessitating measurement at small sample-probe separations. In typical experiments the scanning SET actually produces useful images at separations as large as several mm's, making it very useful to navigate to small devices from afar.

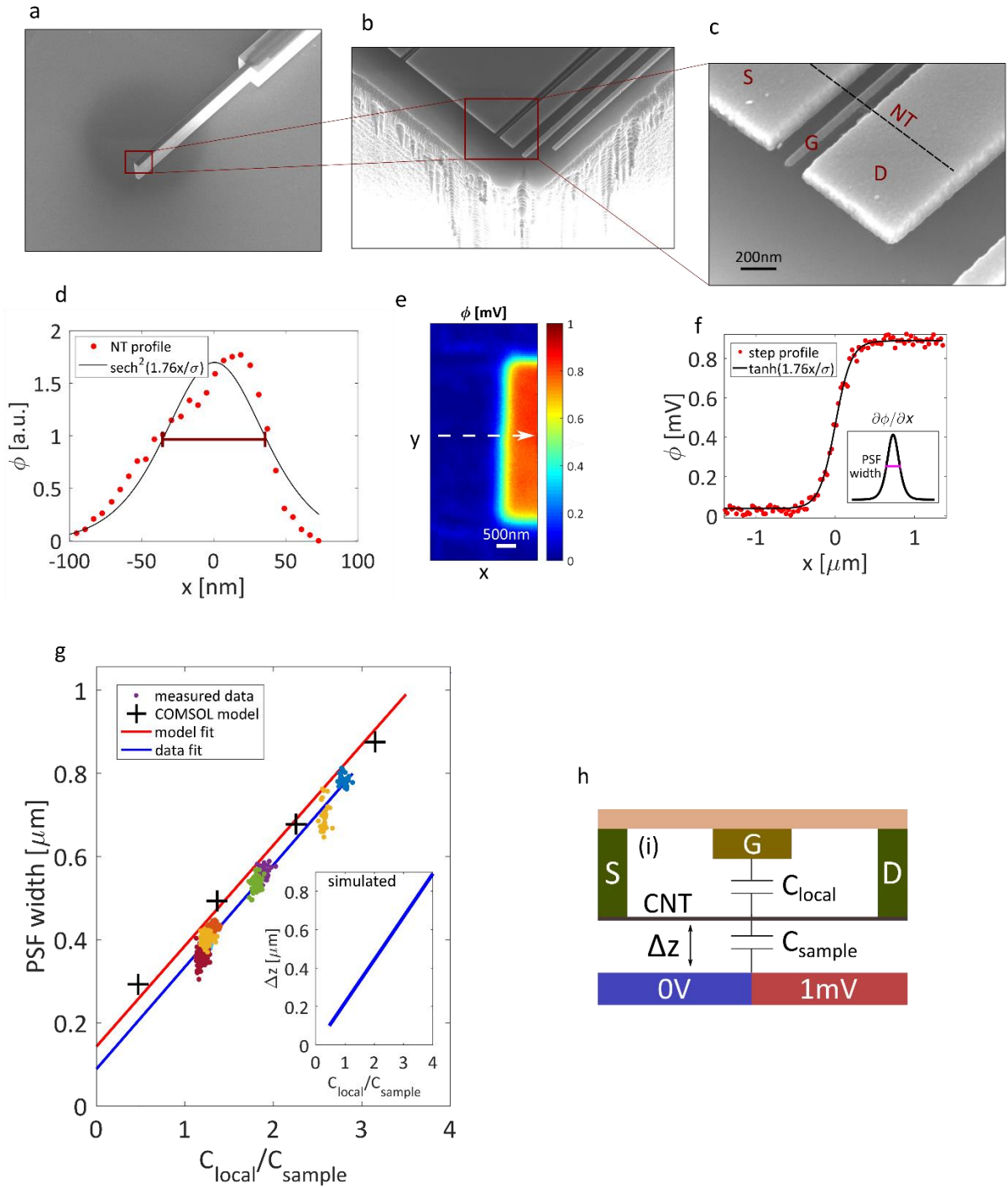

**Figure S1: Determining the spatial resolution. a-c.** SEM images of the scanning SET probe with increasing levels of zoom. In panel **c** The letters 'S', 'D' and 'G' label the source, drain, and gate electrodes, respectively. The nanotube position is marked by the dashed black line. **d.** Determining spatial resolution in 1D by imaging the delta function-

like electrostatic potential of another NT with the SET. The red dots are the measured data, and the black line is a fit to the square of a hyperbolic secant, which matches the data well, giving a FWHM of ~100nm. **e.** SET image of a voltage step in a graphene/hBN device, created by biasing one region (red) with respect to the other (blue). **f.** Determining spatial resolution in 2D from line cut along dashed white arrow in panel e. Red dots are imaged data and the black line is fit to a hyperbolic tangent step function, which matches the data well. Inset: derivative of the step function which gives the PSF. **g**. Measured PSF width as a function of $C_{local}/C_{sample}$, the ratio between the capacitance of the local gate to the SET and the capacitance of the sample to the SET, which is proportional to the height of imaging above the sample $\Delta z$ as shown in the inset. Colored dots represent different data sets, and the blue line is a linear fit. The black crosses are generated from a finite element simulation, which are fit to the red line. **h**. Schematic of the model used in panel g, which includes the source (S), drain (D) and gate (G) metal electrodes of the SET, and the gate's coupling, $C_{local}$, to the nanotube, as well as a split, biased gate with coupling $C_{sample}$ to the nanotube at separation $\Delta z$.

## S2. **Invasiveness**

Scanning probes in general, including the scanning SET have a workfunction that differs from the sample they are imaging. Unless perfectly compensated, this workfunction difference will lead to gating of the sample by the probe, which will change the local carrier density in the sample and likely impact the physics being imaged. This effect, which we refer to as "invasiveness", is particularly pronounced in semiconductors and semimetals whose typical carrier densities are rather low. A typical workfunction difference between probe and sample can be on the order of a few hundred $mV$ and for typical tip-sample capacitances this can easily produce local density changes of the order of $1 \cdot 10^{11} cm^{-2}$. Such invasiveness, can in fact be used as a feature, as is done in the case of scanning gate microscopy[4], which images the response of conductance to strong local disturbance. Our experiments work in the other extreme, where we aim to influence the imaged sample as minimally as possible. This is achieved by making our probe as planar as possible, resulting in probe-sample capacitance that is spatially uniform. This uniform capacitance allows us to null the workfunction difference everywhere by applying a compensating voltage to the probe. We present measurements below that directly quantify our residual invasiveness and demonstrate that it plays a negligible role even in the most sensitive measurements shown in the main text.

To understand the local density variation in the graphene induced by our SET probe, we use the fact that the resistivity of graphene is inversely proportional to its density, $\rho_{xx} \propto n^{-1}$. We choose an etch-defined graphene/hBN device that has two narrow constrictions along the conduction path (Fig. S2a) and record how the 2-point resistance of the device, $R_{2pt}$, varies as the SET is scanned across it. Since transport through this graphene device is dominated by the most resistive constriction, the changes in $R_{2pt}$ are then due primarily to changes in the local density within this constriction induced by the probe. The graphene constriction in this measurement mode essentially then images the SET probe device, whose outline can be seen from figures S2c,d..

The measured dependence of $R_{2pt}$ on back gate voltage (Fig. S2b) displays the usual peak at the charge neutrality point, with additional small oscillations characteristic of the conduction through a constriction at cryogenic temperature. When we set the average density in the graphene to be rather low ($\sim 4 \cdot 10^{10} cm^{-2}$, red point in Fig. S2b) and image the changes in the two point resistance, $\Delta R_{2pt}/R_{2pt}$, due to the scanning SET (Fig. S2c), we observe a maximal change of 2.5%. The corresponding change in carrier density due to probe invasiveness is then $\delta n_{invasiveness} < 1 \cdot 10^9 cm^{-2}$. The density fluctuations in this high-mobility graphene device due to intrinsic disorder, $\delta n_{disorder} \sim 3 \cdot 10^9 cm^{-2}$, are larger than the influence of the SET by a factor of three, verifying that the SET invasiveness is small. In fact, performing the same measurement but at a higher average carrier density in the graphene ($\sim 1.5 \cdot 10^{11} cm^{-2}$, blue dot in Fig. S1b), the imaged $\Delta R_{2pt}/R_{2pt}$ is reduced by an order of magnitude (fig S1d). This is consistent with the claim that $\Delta n/n = -\Delta\rho/\rho$, and shows that at higher carrier densities where ballistic effects are more important, the invasiveness of our probe is even more negligible.

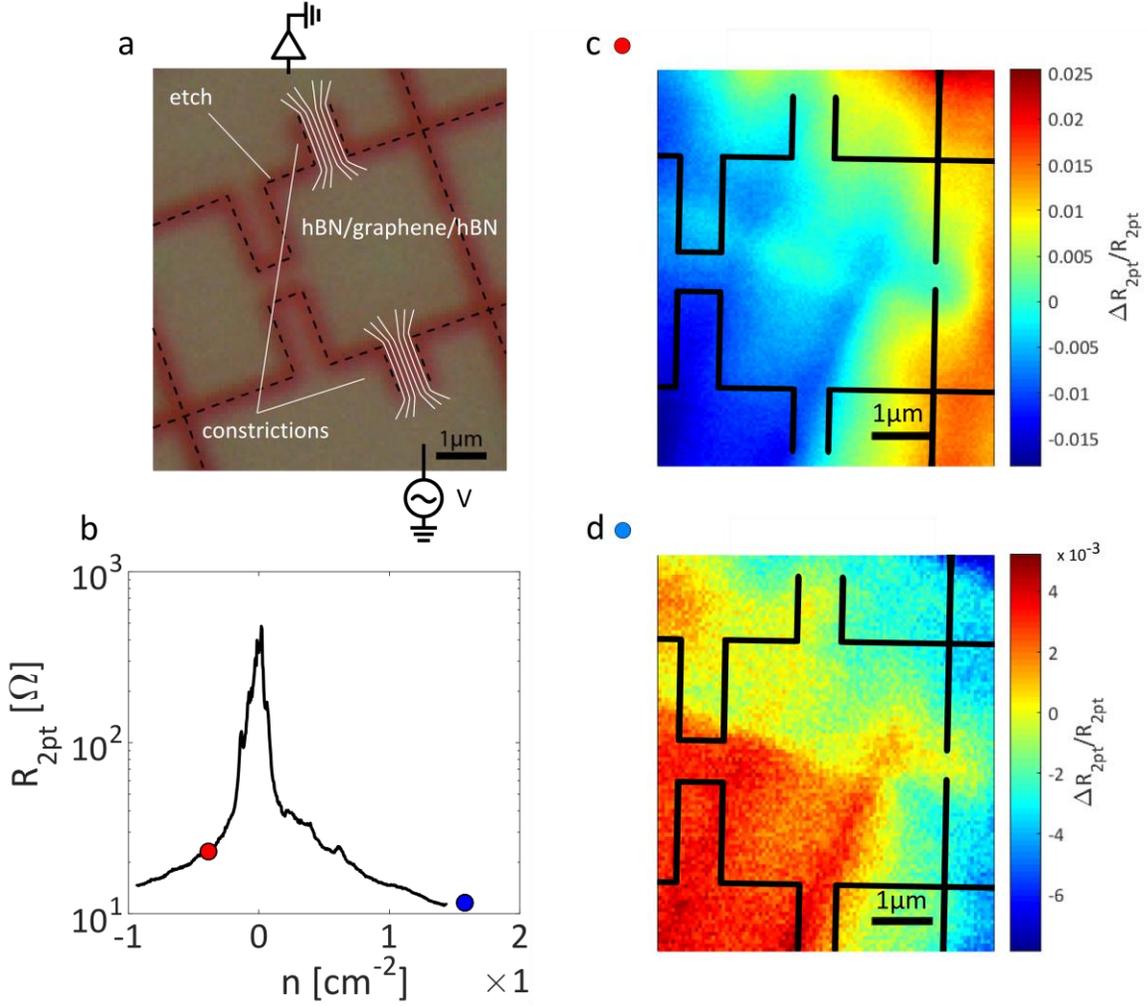

**Figure S2: Invasiveness of the scanning SET. a.** Optical image of the high-mobility Graphene/hBN device used in the invasiveness measurements. The entire field of view in this image has Graphene/hBN, with etched regions marked by black dashed lines. AC voltage is applied between the two labeled constrictions, with white lines indicating current flow. **b**. Measured two-point resistance $R_{2pt}$ between the constrictions in panel a, as a function of the carrier density set by the back gate voltage. **c**. Imaged relative resistance change $\Delta R_{2pt}/R_{2pt}$ as a function of scanning the SET across the graphene device at an average hole density of $\sim 4 \cdot 10^{10} cm^{-2}$ holes (red dot in panel b) The maximum relative resistance change is 2.5%, corresponding to a local density change $\delta n_{invasiveness} < 1 \cdot 10^9 cm^{-2}$, which is less than the density variation due to intrinsic disorder **d**. Imaged relative resistance change at higher (electron) density of $\sim 1.5 \cdot 10^{11} cm^{-2}$, yielding an order of magnitude smaller $\Delta R_{2pt}/R_{2pt}$ than in the previous, already minimally-invasive example in panel c.

## S3. Voltage sensitivity

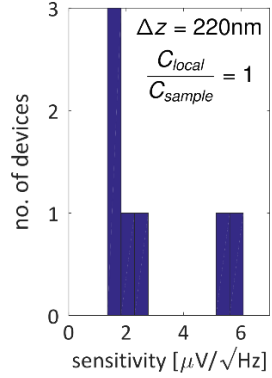

**Figure S3: Voltage sensitivity statistics from multiple scanning NT SET devices**. The measured sensitivities are for a capacitance ratio of 1, as defined in section S1.

The voltage sensitivity of the different SETs used in our measurements is plotted in the histogram in figure S3. The sensitivity is defined as

$$n_V = \sqrt{\langle I_0^2 \rangle} \left(\frac{\partial I_{SET}}{\partial V_{SET}}\right)^{-1} \frac{C_{\text{SET}}}{C_{\text{sample}}} \frac{A}{\sqrt{\Delta f}}$$

where $\sqrt{\langle I_0^2 \rangle}$ is the background measured SET current variance, $\frac{\partial I_{SET}}{\partial V_{SET}}$ is the transconductance of the SET with respect to its local gate (shown schematically in fig 1b inset and in the SEM image in fig S1c above), $\frac{C_{sample}}{C_{local}}$ is a capacitance ratio that quantifies the strength of the voltage coupling between the sample and the SET, $\Delta f$ is the sampling bandwidth, and $A$ is a correction factor to account for the sampling window function. In the histogram above, the noise is plotted at a capacitance ratio of 1, which corresponds to a measurement height of 220nm above the sample under study. We regularly obtain SETs with a voltage sensitivity ~2 µV/√Hz when measuring with audio frequency excitations.

## S4. Imaging electron flow across a resistive obstacle

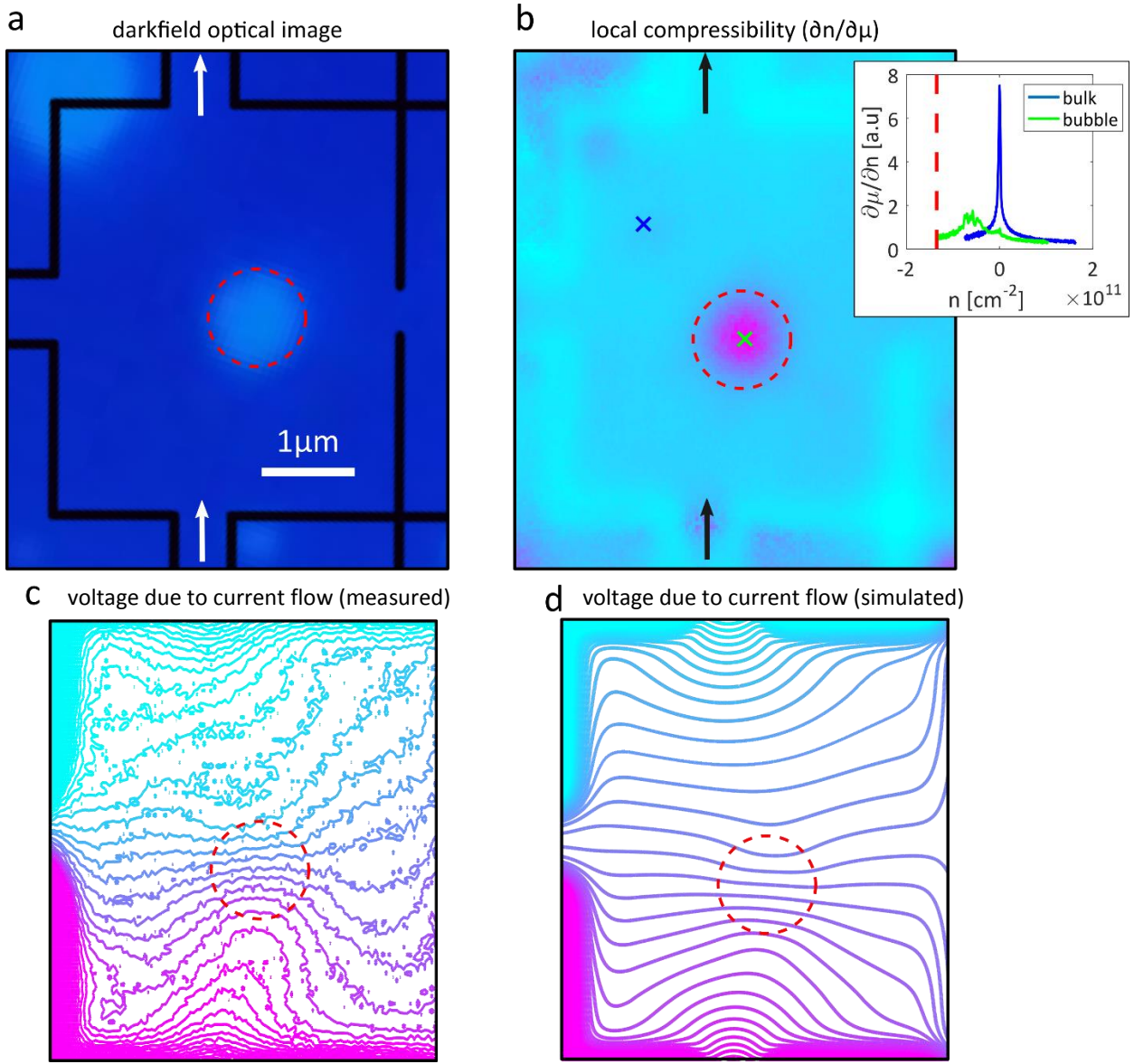

**Figure S4: Imaging electron flow across a resistive obstacle**. **a**. Dark-field optical image of the graphene/hBN (dark blue) device imaged with the SET. The black lines indicate etched boundaries, and the lighter blue regions are hydrocarbon-filled bubbles beneath the graphene formed during fabrication. The central bubble that acts as the resistive obstacle is marked by the red dashed circle, and the white arrows indicate the electron flow path. **b**. Map of the local electronic compressibility at fixed gate voltage measured with the scanning SET. The bright purple spot (marked by red dashed circle) is the bubble, with the black arrows again indicating the current path through the etched constrictions, which are distinguishable in their apparent compressibility from the background. Inset: Inverse compressibility measured at fixed spatial points in the bulk (blue 'x' in main panel) and on the bubble (green 'x' in
9

main panel). As compared to the bulk (blue), the inverse compressibility on the bubble (green) is shifted in density with a broad charge neutrality peak, indicating the graphene in the bubble region is disordered and resistive. The vertical red dashed line marks the density corresponding to the boundary of the bubble region marked by the red circle in the main panel. **c**. Imaged voltage drop of flowing electrons at T=150K. The equipotential contours cluster together near the bubble, marked by the red dashed circle. **d**. Simulation of diffusive flow, where the bubble region (red dashed circle) is taken to have a resistivity 500x that of the surrounding bulk.

We present here an imaged voltage map of electron flow around an obstacle as an additional example of a situation in which conventional transport measurements using fixed probes are incapable of providing a complete understanding of the flow. In the diffusive regime, the voltage drop is clearly related to the local bulk resistivity. By imaging the local voltage drop, we can thus learn about local resistivity variation that would otherwise be characterized in a global, device-scale averaged fashion using transport. A dark field optical image of the device under study is shown in fig S4a. The device is again constructed from a graphene/hBN sandwich (dark blue) with etched boundaries (black lines), but now contains a series of hydrocarbon-filled 'bubbles' (light blue) beneath the graphene that formed spontaneously during device fabrication. These bubbles serve as resistive obstacles to the flow of electrons through the graphene, allowing us to visualize the voltage drop in a highly inhomogeneous device. For the measurements described below, we focus on the influence of the bubble (dashed red circle) located in the center of the etch-defined chamber.

We first locate the position of the bubble within our device by using the SET to measure the local electronic compressibility $\frac{\partial n}{\partial \mu}$ (fig S4b), which is proportional to the density of states and thus allows us to map the location of charge neutrality spatially. The bubble (dashed red circle) is distinguished by its compressibility that differs from the bulk, as are the etched boundaries of the chamber. The inset of S4b shows the inverse compressibility $\frac{\partial \mu}{\partial n}$ as a function of the bulk charge density, taken at the different spatial points indicated in the main panel. Unlike the bulk of the chamber (blue x), the charge neutrality point on the bubble (green x) is shifted in density and diffuse. These measurements indicate that the graphene in the bubble region is comparatively electron-doped and disordered, and can therefore be expected to strongly scatter flowing electrons.

We now turn to a measurement of the voltage drop around the bubble, shown in panel (c). The scan was performed at a bulk density of $-1.4 \times 10^{11} \text{cm}^{-2}$ and at a sample temperature of T=150K to reach the diffusive transport regime (the SET was maintained at T=4K as explained in S5 below), with current entering from the lower constriction of the chamber and exiting through the top constriction. The position of the bubble is again marked with a dashed red circle. As compared to the bulk, the equipotential contours of the flowing electrons bunch together near the bubble and become uniformly spaced, indicating a larger local voltage drop and thus higher resistivity. We emphasize again that this local resistivity feature could not be resolved with a conventional side-contact transport geometry.

At the density and temperature of the scan we expect, from independent measurements, that the bulk mean free path of the graphene will be around 2-3$\mu m$, so that ballistic effects are minimized and the main features of the flow can be captured with a diffusive simulation. In panel (d) we present such a diffusive flow simulation where the bubble has fixed resistivity 500x larger than that of the bulk. The resulting numerical output was then convolved with a point spread function of 150nm (see section S2) in order to compare directly with the measured data. Just as in the data, the simulation shows the same trend of the equipotential lines bunching and becoming uniformly spaced near the bubble. The slight differences between the imaged electrostatic potential and the simulation in the region away from the bubble are likely due to residual ballistic effects.

## S5. Imaging at elevated temperature and in ambient conditions

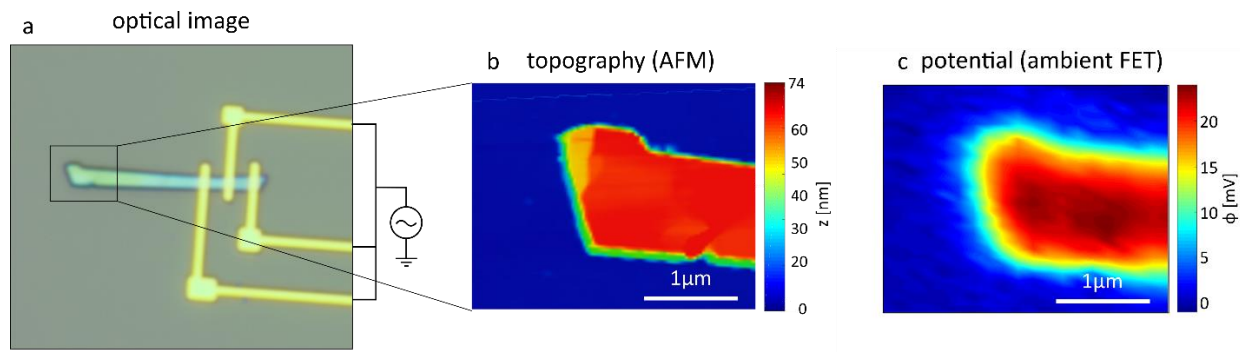

**Figure S5: Imaging in scanning FET mode under ambient conditions**. **a**. Optical image of the studied graphite device, including its electrodes and voltage excitation schematic. The imaged graphite region is marked by the black square. **b**. AFM topograph of the tip of the graphite flake. **c**. Scanning FET voltage map of tip of the graphite under ambient conditions.

Our imaging technique functions at elevated temperatures in two different ways. In the first mode, which was used to obtain the T=150K data presented in section S4 above, we take advantage of a local heater attached to the sample under study. This heater allows us to increase the temperature of the sample while the SET, which is strongly thermally coupled to the helium bath in our microscope, is maintained at T=4K. Using this approach, the sample can in fact be heated to room temperature while the SET is fixed at cryogenic temperature, despite only ~100nm separation between the two microchips. Overall we find that both the voltage sensitivity of the SET and the vibrations/stability of the measurements are largely unaffected by varying the temperature.

In the second mode of measuring at elevated temperature, we operate the nanotube device not as an SET, but instead as a scanning field effect transistor (FET). In this mode, the scanning FET can operate at ambient conditions (room temperature, no vacuum), using the intrinsic bandgap of the semiconducting nanotube segment to transduce the local electrostatic potential of the sample under study into a measureable current. Just as in the SET mode, no current directly passes between the sample and the FET, allowing us to similarly image non-invasively. As an example of imaging with the NT FET, we scanned a simple representative device consisting of a graphite flake to which a voltage excitation is applied, while maintaining a grounded back gate. An optical image of the device is shown in fig S5a, including its metallic contacts and a schematic of the voltage excitation. In fig S5b, we present AFM characterization of the topography of the end of the graphite section (outlined in black square in S5a). We then image the device with the FET under ambient conditions (fig S5c), noting that we are now recording the transconductance $\partial I_{FET}/\partial V_{sample}$ of the FET at each spatial point, rather than single electron tunneling of a coulomb-blockaded SET. The FET image of the graphite device matches the AFM image closely, though with a spatial resolution of $85 nm$, limited by the height of scanning. The ultimate spatial resolution of this technique (few tens of nm) is identical to the limit when imaging as an SET (section S1). The voltage sensitivity at ambient conditions of the FET is $\sim 10 \mu V/\sqrt{Hz}$, which compares favorably to the performance when operated as an SET.